\documentclass[aps,prl,twocolumn,floatfix]{revtex4}
\usepackage{graphicx,units,xspace,amsmath}

\graphicspath{{figures/}}

\newcommand{\micron}{\ensuremath{\unit{\mu m}}\xspace}

\begin{document}

\title{Observation of Flux Reversal in a Symmetric Optical Thermal Ratchet}

\author{Sang-Hyuk Lee}

\author{Kosta Ladavac}
\altaffiliation{Dept.\ of Physics, James Franck Institute and
Institute for Biophysical Dynamics,
The University of Chicago, Chicago, IL 60637}

\author{Marco Polin}

\author{David G. Grier}

\affiliation{Department of Physics and Center for Soft Matter
  Research\\
New York University, New York, NY 10003}

\date{\today}

\begin{abstract}
  We demonstrate that a cycle of three holographic optical trapping patterns
  can implement a thermal ratchet for diffusing colloidal 
  spheres, and that the ratchet-driven transport displays
  flux reversal as a function of the cycle frequency and the inter-trap separation.
  Unlike previously described ratchet models, the approach we describe
  involves three equivalent states, each of which is locally and globally 
  spatially symmetric, with spatiotemporal symmetry being broken
  by the sequence of states.
\end{abstract}

\maketitle

Brownian motion cannot create a steady flux
in a system at equilibrium.
Nor can local asymmetries in a static potential energy landscape rectify
Brownian motion to induce a drift.
A landscape that varies in time, however, can eke a flux out of
random fluctuations by
breaking spatiotemporal symmetry 
\cite{astumian94,prost94,chauwin94,rousselet94}.
Such flux-inducing time-dependent potentials are known as thermal ratchets
\cite{reimann02,linke02}, and their ability to bias diffusion
by rectifying thermal fluctuations
has been proposed as a possible mechanism for transport by molecular motors
and is being actively exploited for macromolecular sorting \cite{hughes02}.

Most thermal ratchet models are based on spatially asymmetric potentials.
Their time variation involves displacing or tilting them
relative to the laboratory frame, modulating
their amplitude, changing their periodicity, or some combination,
usually in a two-state cycle.  
Chen demonstrated that a
spatially symmetric potential still can induce drift, provided that it
is applied in a three-state cycle, one of which
allows for free diffusion \cite{chen97}.
This idea since has been refined \cite{kananda99}
and generalized \cite{savelev02}.

The space-filling
potential energy landscapes required for most
such models pose technical challenges.
Furthermore, their
relationship to the operation of natural thermal ratchets has proved
difficult to establish.

This Letter describes the first experimental demonstration of 
a spatially symmetric thermal ratchet, which we have implemented with
holographic optical traps \cite{dufresne98,dufresne01a,curtis02}.
The potential energy landscape in this system consists of a large
number of 
discrete optical tweezers \cite{ashkin86}, each of which acts as
a symmetric potential energy well for nanometer- to micrometer-scale objects
such as colloidal spheres.
We arrange these wells so
that colloidal spheres can diffuse freely in the interstitial spaces
but are localized rapidly once they encounter a trap.
A three-state thermal ratchet then requires only displaced 
copies of a single two-dimensional trapping pattern.
Despite its simplicity, this ratchet model displays flux reversal
\cite{bier96,reimann02}
in which the direction of motion is controlled by a
balance between the rate at which particles diffuse across the
landscape and the ratchet's cycling rate.

\begin{figure}[tbh]
  \centering
  \includegraphics[width=.75\columnwidth]{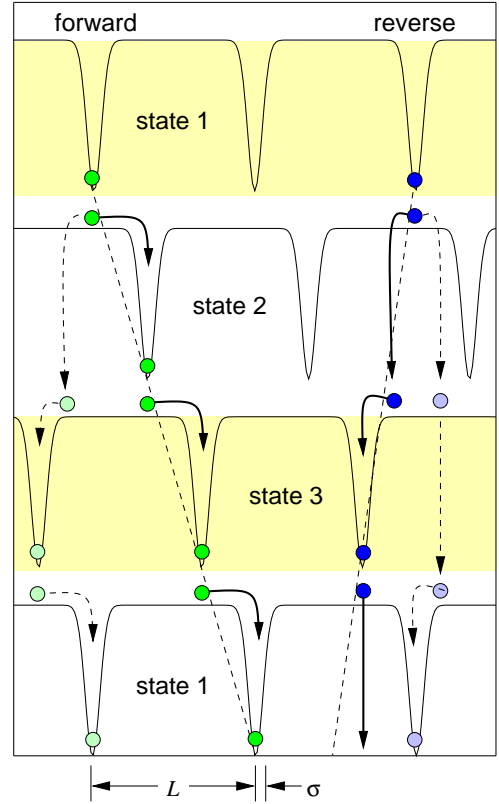}
  \caption{A spatially-symmetric three-state ratchet potential
    comprised of discrete potential wells.}
  \label{fig:peristalsis}
\end{figure}

Often predicted, and inferred from the behavior of 
some natural molecular motors and semiconductor devices
\cite{reimann02}, 
flux reversal has been directly observed in comparatively few
systems.
Previous demonstrations have focused on
ratcheting of magnetic flux quanta through type-II superconductors
in both the quantum mechanical \cite{carapella02} 
and classical \cite{villegas03} regimes, or else have exploited 
the crossover from quantum mechanical to classical transport in a quantum dot
array \cite{linke99}.
Unlike the present implementation, these exploit
spatially asymmetric potentials and take the form of
rocking ratchets \cite{reimann02}.
A massively parallel 
hydrodynamic ratchet for fluid-borne objects driven
by oscillatory flows
through asymmetric pores also shows 
size-dependent flux reversal \cite{kettner00,matthias03}.
In this case, however, the force field is provided by the
divergence-free flow of an incompressible fluid and so cannot be described
as arising from a potential energy landscape.
Rather, this is an instance of a so-called drift ratchet \cite{kettner00}.
Other pioneering implementations of classical force-free
thermal ratchets also were based on asymmetric potentials,
but did not exhibit flux reversal
\cite{faucheux95,faucheux95a,gorretalini98,bader99,bader02}.

Figure~\ref{fig:peristalsis} shows the principle upon which the
three-state optical thermal ratchet operates.
The process starts out with a pattern of discrete optical traps,
each of which can localize an object.
The pattern in the initial state is schematically represented as three discrete 
potential energy wells, each of width $\sigma$ and depth $V_0$, 
separated by distance $L$.
A practical trapping pattern can include a great many optical traps organized
into manifolds.
The first pattern is extinguished after time $T$ and replaced 
immediately with the
second, which is displaced from the first by $L/3$. 
This is repeated in the third state with an additional step of $L/3$,
and again when the cycle is
completed by returning to the first state

If the traps in a given state overlap those
in the state before, a trapped particle is transported
deterministically forward.
Running through this cycle repeatedly transfers the object 
in a direction
determined unambiguously by the sequence of states, and is known as
optical peristalsis \cite{koss03}.
The direction of motion can be reversed only by reversing the 
sequence.

The optical thermal ratchet differs from this in that the 
inter-trap separation $L$
is substantially larger than $\sigma$.
Consequently, particles trapped in the first pattern are released into
a force-free region and can diffuse freely when
that pattern is replaced by the second.
Those particles, such as the example labeled ``forward''
in Fig.~\ref{fig:peristalsis},
that diffuse far enough to reach the nearest traps 
in the second pattern rapidly become localized.
A comparable proportion of this localized fraction then can be transferred 
forward again once the
third pattern is projected, 
and again when the cycle returns to the first state.

Unlike optical peristalsis, in which all particles are promoted in each cycle,
the stochastic ratchet transfers only a fraction.
This, however, leads to a new opportunity.
Other particles
that are too slow to catch the forward-going wave might still reach
a trap on the opposite side of
their starting point while the third pattern is illuminated.
These particles would be transferred \emph{backward} 
by $L/3$ after time $2T$, as shown in the ``reverse'' trajectory
in Fig.~\ref{fig:peristalsis}.

For particles of diffusivity $D$, the time required to diffuse
the inter-trap separation is $\tau = L^2/(2D)$.
If we assume that particles begin each cycle well localized at
a trap, and that the traps are well separated compared to their
widths, then the probability for ratcheting forward by $L/3$
during the interval $T$ is
roughly $P_F \approx \exp(-(L/3)^2 / (2DT))$, while the probability
of ratcheting backwards in time $2T$ is roughly
$P_R \approx \exp(-(L/3)^2 / (4DT))$.  The associated fluxes of
particles then are $v_F = P_F L / (3T)$ and $v_R = -P_R L / (6T)$,
with the dominant term determining the overall direction of motion.
Crudely, then, we expect the direction of induced motion to
reverse when $T / \tau \lesssim (18 \ln 2)^{-1} \approx 0.08$.

More formally, we can 
model an array of evenly spaced optical traps
in the $n$-th pattern as Gaussian potential wells
\begin{equation}
  \label{eq:v}
  V_n(x) = \sum_{j = -N}^N -V_0 \, \exp \left(
    - \frac{\left(x - jL - n \,\frac{L}{3}\right)^2}{2 \sigma^2} \right),
\end{equation}
where $n = 0$, 1, or 2, and $N$ sets the extent of the landscape.
The probability density $\rho(x,t) \, dx$ for finding a Brownian particle 
within $dx$ of position $x$ at time $t$ in state $n$ evolves according to the
master equation \cite{risken89}
\begin{equation}
  \label{eq:master}
  \rho(y, t + T) = \int P_n(y, T| x, 0) \, \rho(x,t) \, dx,
\end{equation}
characterized by the propagator
\begin{equation}
  \label{eq:propagator}
  P_n(y, T | x, 0) = e^{L_n(y) \, T} \, \delta(y-x),
\end{equation}
where the Liouville operator for state $n$ is
\begin{equation}
  L_n(y) = D \, \left(\frac{\partial^2}{\partial y^2} - 
    \beta \frac{\partial}{\partial y} V^\prime_n(y) \right),
\end{equation}
with $V_n^\prime(y) = \frac{dV_n}{dy}$,
and where $\beta^{-1}$ is the thermal energy scale.

The master equation for a three-state cycle is
\begin{equation}
  \label{eq:master3}
  \rho(y,t+3T) = \int P_{123}(y,3T|x,0) \, \rho(x,t) \, dx,
\end{equation}
with the three-state propagator
\begin{multline}
  P_{123}(y,3T|x,0) = \int dy_1 \, dy_2 \, P_3(y,T|y_2,0) \times \\
  P_2(y_2,T|y_1,0) \, P_1(y_1,T|x,0).
\end{multline}
Because the landscape is periodic and analytic,
Eq.~(\ref{eq:master3}) has a
steady-state solution such that
\begin{align}
  \rho(x,t+3T) & = \rho(x,t) \\
  & \equiv \rho_{123}(x).
\end{align}
The mean velocity of this steady-state then is given by
\begin{equation}
  v = \int P_{123}(y,3T|x,0) \, \left(\frac{y-x}{3T}\right) \, \rho_{123}(x) \, dx \, dy,
  \label{eq:velocity}
\end{equation}
where $P_{123}(y,3T|x,0)$ is
the probability for a particle originally
at position $x$ to ``jump'' to position $y$ by the end of one compete cycle,
$(y-x)/(3T)$ is the velocity associated with making such a jump, and $\rho_{123}(x)$
is the fraction of the available particles actually at $x$ at the beginning of
the cycle in steady-state.
This formulation is invariant with respect to cyclic permutations of the states, so that the
same flux of particles would be measured at the end of each state.
The average velocity $v$ therefore describes the time-averaged flux of particles
driven by the ratchet.

\begin{figure}[htbp]
  \centering
  \includegraphics[width=\columnwidth]{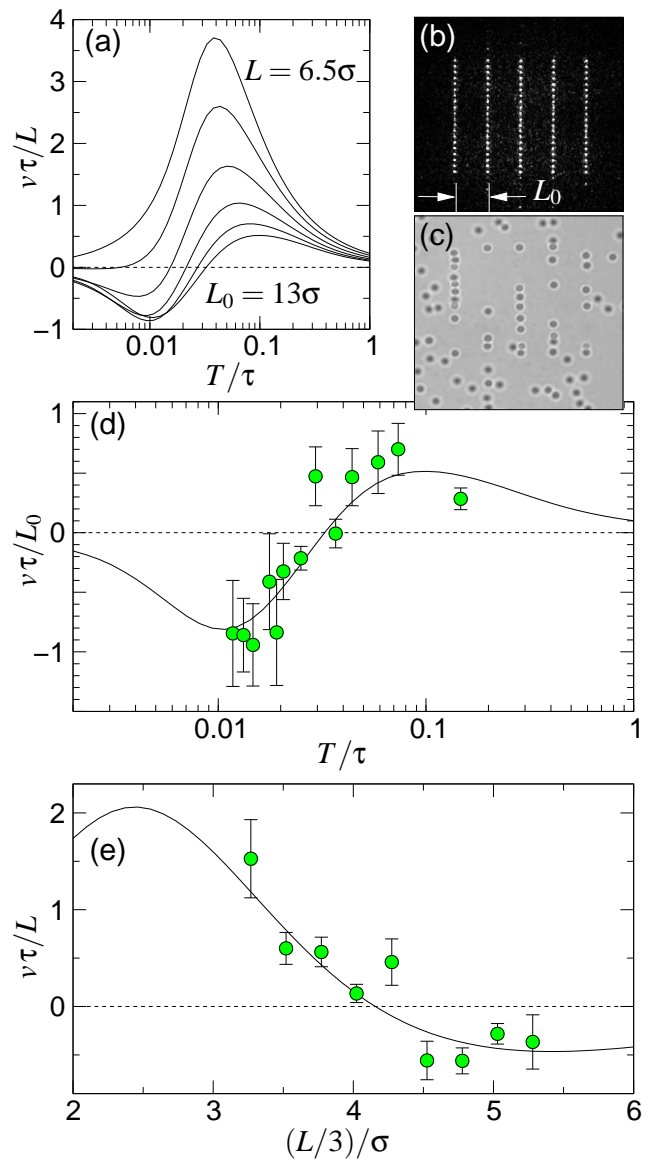}
  \caption{(a) Crossover from deterministic optical peristalsis at $L = 6.5~\sigma$
    to thermal ratchet behavior with flux reversal at $L = 13~\sigma$ for a three-state
    cycle of Gaussian well potentials at $\beta V_0 = 8.5$, $\sigma = 0.53~\micron$
    and $D = 0.33~\unit{\micron^2/sec}$.
    Intermediate curves are calculated for evenly spaced values of $L$.
    (b) Image of $20 \times 5$ array of holographic optical traps at $L_0 = 6.7~\micron$.
    (c) Image of colloidal silica spheres 1.53~\micron in diameter interacting with the array.
    (d) Rate dependence of the induced drift velocity for fixed inter-trap separation, $L_0$.
    (e) Separation dependence for fixed inter-state delay, $T = 2~\unit{sec}$.
  }
  \label{fig:fluxreversal}
\end{figure}

Figure \ref{fig:fluxreversal}(a) shows numerical solutions of this system of equations for
representative values of the relative inter-well separation $L/\sigma$.
If the interval $T$ between states is very short,
particles are unable to keep up with
the evolving potential energy landscape, and so never travel far from
their initial positions; the mean velocity
vanishes in this limit.
The transport speed $v$ also vanishes as $1/T$ for large values of $T$
because the induced drift becomes
limited by the delay between states.
If traps in consecutive patterns are close enough ($L = 6.5~\sigma$ in Fig.~\ref{fig:fluxreversal}(a)) 
particles jump forward at each transition with high probability, 
yielding a uniformly positive drift velocity.
This transfer reaches its maximum efficiency for moderate cycle times,
$T/\tau \approx 2 \sqrt{2} (L/\sigma) (\beta V_0)^{-1}$.
More widely separated traps ($L = 13~\sigma$ in Fig.~\ref{fig:fluxreversal}(a)) yield more
interesting behavior.
Here, particles are able to keep up with the forward-going wave for large values
of $T$.  Faster cycling, however, leads to flux reversal,
characterized by 
negative values of $v$.

We implemented this thermal ratchet protocol
for a sample of 1.53~\micron diameter colloidal
silica spheres (Bangs Laboratories, lot number 5328)
dispersed in water, using potential energy landscapes
created from arrays of holographic optical traps
\cite{dufresne98,dufresne01a,curtis02}.
The sample was enclosed in a hermetically sealed glass chamber
roughly 40~\micron thick
created by bonding the edges of a coverslip to a microscope slide
and was allowed to equilibrate to room temperature ($21 \pm 1^\circ\unit{C}$)
on the stage of a Zeiss S100TV Axiovert inverted optical microscope.
A $100\times$ NA 1.4 oil immersion SPlan Apo objective lens was used to
focus the optical tweezer array into the sample and to image the
spheres,
whose motions were captured with
an NEC TI 324A low noise monochrome CCD camera.
The micrograph
in Fig.~\ref{fig:fluxreversal}(b)
shows the focused light from a $20 \times 5$ array of optical traps
formed by a phase hologram projected with a Hamamatsu X7550 
spatial light modulator \cite{igasaki99}.
The tweezers are arranged in twenty-trap manifolds $25~\micron$ long 
separated by $L_0 = 6.7~\micron$.
Each trap is powered by an estimated $2.5 \pm 0.4~\unit{mW}$ 
of laser light at 532~\unit{nm}.
The particles, which appear in the bright-field micrograph in Fig.~\ref{fig:fluxreversal}(c),
are twice as dense as water and sediment to the lower glass surface,
where they diffuse freely in the plane with a measured
diffusion coefficient
of $D = 0.33 \pm 0.03~\unit{\micron^2/sec}$, which reflects the
influence of the nearby wall.
Out-of-plane fluctuations were minimized by projecting the traps at the
spheres' equilibrium height above the wall \cite{behrens03}.

We projected three-state cycles of optical trapping patterns in which the
manifolds in Fig.~\ref{fig:fluxreversal}(b)
were displaced horizontally by $-L_0/3$, 0, and $L_0/3$, 
with inter-state delay times $T$ ranging from
0.8~\unit{sec} to 10~\unit{sec}.
The particles' motions
were recorded as uncompressed digital video streams for
analysis \cite{crocker96}.
Between 40 and 60 particles were in the trapping pattern during 
a typical run, so that roughly 40 cycles sufficed to acquire reasonable statistics 
under each set of conditions
without complications due to collisions.
We also tracked particles outside the trapping pattern to monitor their diffusion coefficients
and to ensure the absence of drifts in the supporting fluid.
The results plotted in Fig.~\ref{fig:fluxreversal}(d) reveal flux
reversal at $T / \tau \approx 0.03$.
Excellent agreement with
Eq.~(\ref{eq:velocity}) is obtained for
$\beta V_0 = 8.5 \pm 0.8$ and 
$\sigma = 0.53 \pm 0.01~\micron$.

The appearance of flux reversal as one parameter is varied implies
that other parameters also should control the direction of motion \cite{reimann02}.
Indeed, flux reversal is obtained 
in Fig.~\ref{fig:fluxreversal}(e)
as the inter-trap separation is varied from $L = 5.1~\micron$ to 8.3~\micron
at fixed delay time, $T = 2~\sec$.
These results also agree well with predictions of
Eq.~(\ref{eq:velocity}),
with no adjustable parameters.
The same effect also should arise
for different populations in a heterogeneous sample with
different values of $D$, $V_0$ and $\sigma$ \cite{ladavac04,pelton04a}.
In this case, distinct fractions
can be induced to move simultaneously in opposite directions.

Such sensitivity of the transport direction to details of
the dynamics also might play a role in the functioning of molecular
motors such as myosin-VI whose retrograde motion on actin
filaments compared with other myosins 
has excited much interest \cite{hasson01}.
This molecular motor is known to be nonprocessive \cite{lister04}; 
its motion involves a diffusive
search of the actin filament's potential energy landscape,
which nevertheless results in unidirectional hand-over-hand
transport \cite{okten04}.  These characteristics are consistent with
the present model's timing-based flux reversal mechanism, and could
provide a basis to explain how small structural differences among myosins
could lead to oppositely directed transport.

Brian Koss contributed to the early stages of this project.  We also
have benefited from
numerous conversations with Dean Astumian and Martin Bier, and more recently with
Franco Nori and Stephen Quake.  We also are grateful for Mark Ofitserov's contributions to the optical
trapping system.
This work was supported by the National Science Foundation through Grant Number
DBI-0233971 and Grant Number DMR-0304906.
S.-H. Lee acknowledges support from a Kessler Family Foundation Fellowship.


\begin{thebibliography}{35}
\expandafter\ifx\csname natexlab\endcsname\relax\def\natexlab#1{#1}\fi
\expandafter\ifx\csname bibnamefont\endcsname\relax
  \def\bibnamefont#1{#1}\fi
\expandafter\ifx\csname bibfnamefont\endcsname\relax
  \def\bibfnamefont#1{#1}\fi
\expandafter\ifx\csname citenamefont\endcsname\relax
  \def\citenamefont#1{#1}\fi
\expandafter\ifx\csname url\endcsname\relax
  \def\url#1{\texttt{#1}}\fi
\expandafter\ifx\csname urlprefix\endcsname\relax\def\urlprefix{URL }\fi
\providecommand{\bibinfo}[2]{#2}
\providecommand{\eprint}[2][]{\url{#2}}

\bibitem[{\citenamefont{Astumian and Bier}(1994)}]{astumian94}
\bibinfo{author}{\bibfnamefont{R.~D.} \bibnamefont{Astumian}} \bibnamefont{and}
  \bibinfo{author}{\bibfnamefont{M.}~\bibnamefont{Bier}},
  \bibinfo{journal}{Phys. Rev. Lett.} \textbf{\bibinfo{volume}{72}},
  \bibinfo{pages}{1766} (\bibinfo{year}{1994}).

\bibitem[{\citenamefont{Prost et~al.}(1994)\citenamefont{Prost, Chauwin,
  Peliti, and Ajdari}}]{prost94}
\bibinfo{author}{\bibfnamefont{J.}~\bibnamefont{Prost}},
  \bibinfo{author}{\bibfnamefont{J.~F.} \bibnamefont{Chauwin}},
  \bibinfo{author}{\bibfnamefont{L.}~\bibnamefont{Peliti}}, \bibnamefont{and}
  \bibinfo{author}{\bibfnamefont{A.}~\bibnamefont{Ajdari}},
  \bibinfo{journal}{Phys. Rev. Lett.} \textbf{\bibinfo{volume}{72}},
  \bibinfo{pages}{2652} (\bibinfo{year}{1994}).

\bibitem[{\citenamefont{Chauwin et~al.}(1994)\citenamefont{Chauwin, Ajdari, and
  Prost}}]{chauwin94}
\bibinfo{author}{\bibfnamefont{J.~F.} \bibnamefont{Chauwin}},
  \bibinfo{author}{\bibfnamefont{A.}~\bibnamefont{Ajdari}}, \bibnamefont{and}
  \bibinfo{author}{\bibfnamefont{J.}~\bibnamefont{Prost}},
  \bibinfo{journal}{Europhys. Lett.} \textbf{\bibinfo{volume}{27}},
  \bibinfo{pages}{421} (\bibinfo{year}{1994}).

\bibitem[{\citenamefont{Rousselet et~al.}(1994)\citenamefont{Rousselet, Salome,
  Ajdari, and Prost}}]{rousselet94}
\bibinfo{author}{\bibfnamefont{J.}~\bibnamefont{Rousselet}},
  \bibinfo{author}{\bibfnamefont{L.}~\bibnamefont{Salome}},
  \bibinfo{author}{\bibfnamefont{A.}~\bibnamefont{Ajdari}}, \bibnamefont{and}
  \bibinfo{author}{\bibfnamefont{J.}~\bibnamefont{Prost}},
  \bibinfo{journal}{Nature} \textbf{\bibinfo{volume}{370}},
  \bibinfo{pages}{446} (\bibinfo{year}{1994}).

\bibitem[{\citenamefont{Reimann}(2002)}]{reimann02}
\bibinfo{author}{\bibfnamefont{P.}~\bibnamefont{Reimann}},
  \bibinfo{journal}{Phys. Rep.} \textbf{\bibinfo{volume}{361}},
  \bibinfo{pages}{57} (\bibinfo{year}{2002}).

\bibitem[{\citenamefont{Linke}(2002)}]{linke02}
\bibinfo{author}{\bibfnamefont{H.}~\bibnamefont{Linke}},
  \bibinfo{journal}{Appl. Phys. A} \textbf{\bibinfo{volume}{75}},
  \bibinfo{pages}{167} (\bibinfo{year}{2002}).

\bibitem[{\citenamefont{Hughes}(2002)}]{hughes02}
\bibinfo{author}{\bibfnamefont{M.~P.} \bibnamefont{Hughes}},
  \bibinfo{journal}{Electrophoresis} \textbf{\bibinfo{volume}{23}},
  \bibinfo{pages}{2569} (\bibinfo{year}{2002}).

\bibitem[{\citenamefont{Chen}(1997)}]{chen97}
\bibinfo{author}{\bibfnamefont{Y.-d.} \bibnamefont{Chen}},
  \bibinfo{journal}{Phys. Rev. Lett.} \textbf{\bibinfo{volume}{79}},
  \bibinfo{pages}{3117} (\bibinfo{year}{1997}).

\bibitem[{\citenamefont{Kananda and Sasaki}(1999)}]{kananda99}
\bibinfo{author}{\bibfnamefont{R.}~\bibnamefont{Kananda}} \bibnamefont{and}
  \bibinfo{author}{\bibfnamefont{K.}~\bibnamefont{Sasaki}},
  \bibinfo{journal}{J. Phys. Soc. Japan}
  \textbf{\bibinfo{volume}{68}}, \bibinfo{pages}{3759} (\bibinfo{year}{1999}).

\bibitem[{\citenamefont{Savel'ev and Nori}(2002)}]{savelev02}
\bibinfo{author}{\bibfnamefont{S.}~\bibnamefont{Savel'ev}} \bibnamefont{and}
  \bibinfo{author}{\bibfnamefont{F.}~\bibnamefont{Nori}},
  \bibinfo{journal}{Nature Materials} \textbf{\bibinfo{volume}{1}},
  \bibinfo{pages}{179} (\bibinfo{year}{2002}).

\bibitem[{\citenamefont{Dufresne and Grier}(1998)}]{dufresne98}
\bibinfo{author}{\bibfnamefont{E.~R.} \bibnamefont{Dufresne}} \bibnamefont{and}
  \bibinfo{author}{\bibfnamefont{D.~G.} \bibnamefont{Grier}},
  \bibinfo{journal}{Rev. Sci. Instr.} \textbf{\bibinfo{volume}{69}},
  \bibinfo{pages}{1974} (\bibinfo{year}{1998}).

\bibitem[{\citenamefont{Dufresne et~al.}(2001)\citenamefont{Dufresne, Spalding,
  Dearing, Sheets, and Grier}}]{dufresne01a}
\bibinfo{author}{\bibfnamefont{E.~R.} \bibnamefont{Dufresne}},
  \bibinfo{author}{\bibfnamefont{G.~C.} \bibnamefont{Spalding}},
  \bibinfo{author}{\bibfnamefont{M.~T.} \bibnamefont{Dearing}},
  \bibinfo{author}{\bibfnamefont{S.~A.} \bibnamefont{Sheets}},
  \bibnamefont{and} \bibinfo{author}{\bibfnamefont{D.~G.} \bibnamefont{Grier}},
  \bibinfo{journal}{Rev. Sci. Instr.} \textbf{\bibinfo{volume}{72}},
  \bibinfo{pages}{1810} (\bibinfo{year}{2001}).

\bibitem[{\citenamefont{Curtis et~al.}(2002)\citenamefont{Curtis, Koss, and
  Grier}}]{curtis02}
\bibinfo{author}{\bibfnamefont{J.~E.} \bibnamefont{Curtis}},
  \bibinfo{author}{\bibfnamefont{B.~A.} \bibnamefont{Koss}}, \bibnamefont{and}
  \bibinfo{author}{\bibfnamefont{D.~G.} \bibnamefont{Grier}},
  \bibinfo{journal}{Opt. Comm.} \textbf{\bibinfo{volume}{207}},
  \bibinfo{pages}{169} (\bibinfo{year}{2002}).

\bibitem[{\citenamefont{Ashkin et~al.}(1986)\citenamefont{Ashkin, Dziedzic,
  Bjorkholm, and Chu}}]{ashkin86}
\bibinfo{author}{\bibfnamefont{A.}~\bibnamefont{Ashkin}},
  \bibinfo{author}{\bibfnamefont{J.~M.} \bibnamefont{Dziedzic}},
  \bibinfo{author}{\bibfnamefont{J.~E.} \bibnamefont{Bjorkholm}},
  \bibnamefont{and} \bibinfo{author}{\bibfnamefont{S.}~\bibnamefont{Chu}},
  \bibinfo{journal}{Opt. Lett.} \textbf{\bibinfo{volume}{11}},
  \bibinfo{pages}{288} (\bibinfo{year}{1986}).

\bibitem[{\citenamefont{Bier and Astumian}(1996)}]{bier96}
\bibinfo{author}{\bibfnamefont{M.}~\bibnamefont{Bier}} \bibnamefont{and}
  \bibinfo{author}{\bibfnamefont{R.~D.} \bibnamefont{Astumian}},
  \bibinfo{journal}{Phys. Rev. Lett.} \textbf{\bibinfo{volume}{76}},
  \bibinfo{pages}{4277} (\bibinfo{year}{1996}).

\bibitem[{\citenamefont{Carapella et~al.}(2002)\citenamefont{Carapella,
  Costabile, Martucciello, Cirillo, Latempa, Polcari, and
  Filatrella}}]{carapella02}
\bibinfo{author}{\bibfnamefont{G.}~\bibnamefont{Carapella}},
  \bibinfo{author}{\bibfnamefont{G.}~\bibnamefont{Costabile}},
  \bibinfo{author}{\bibfnamefont{N.}~\bibnamefont{Martucciello}},
  \bibinfo{author}{\bibfnamefont{M.}~\bibnamefont{Cirillo}},
  \bibinfo{author}{\bibfnamefont{R.}~\bibnamefont{Latempa}},
  \bibinfo{author}{\bibfnamefont{A.}~\bibnamefont{Polcari}}, \bibnamefont{and}
  \bibinfo{author}{\bibfnamefont{G.}~\bibnamefont{Filatrella}},
  \bibinfo{journal}{Physica C} \textbf{\bibinfo{volume}{382}},
  \bibinfo{pages}{337} (\bibinfo{year}{2002}).

\bibitem[{\citenamefont{Villegas et~al.}(2003)\citenamefont{Villegas, Savel'ev,
  Nori, Gonzalez, Anguita, Garcia, and Vicent}}]{villegas03}
\bibinfo{author}{\bibfnamefont{J.~E.} \bibnamefont{Villegas}},
  \bibinfo{author}{\bibfnamefont{S.}~\bibnamefont{Savel'ev}},
  \bibinfo{author}{\bibfnamefont{F.}~\bibnamefont{Nori}},
  \bibinfo{author}{\bibfnamefont{E.~M.} \bibnamefont{Gonzalez}},
  \bibinfo{author}{\bibfnamefont{J.~V.} \bibnamefont{Anguita}},
  \bibinfo{author}{\bibfnamefont{R.}~\bibnamefont{Garcia}}, \bibnamefont{and}
  \bibinfo{author}{\bibfnamefont{J.~L.} \bibnamefont{Vicent}},
  \bibinfo{journal}{Science} \textbf{\bibinfo{volume}{302}},
  \bibinfo{pages}{1188} (\bibinfo{year}{2003}).

\bibitem[{\citenamefont{Linke et~al.}(1999)\citenamefont{Linke, Humphrey,
  L\"{o}fgren, Sushkov, Newbury, Taylor, and Omling}}]{linke99}
\bibinfo{author}{\bibfnamefont{H.}~\bibnamefont{Linke}},
  \bibinfo{author}{\bibfnamefont{T.~E.} \bibnamefont{Humphrey}},
  \bibinfo{author}{\bibfnamefont{A.}~\bibnamefont{L\"{o}fgren}},
  \bibinfo{author}{\bibfnamefont{A.~O.} \bibnamefont{Sushkov}},
  \bibinfo{author}{\bibfnamefont{R.}~\bibnamefont{Newbury}},
  \bibinfo{author}{\bibfnamefont{R.~P.} \bibnamefont{Taylor}},
  \bibnamefont{and} \bibinfo{author}{\bibfnamefont{P.}~\bibnamefont{Omling}},
  \bibinfo{journal}{Science} \textbf{\bibinfo{volume}{286}},
  \bibinfo{pages}{2314} (\bibinfo{year}{1999}).

\bibitem[{\citenamefont{Matthias and M\"{u}ller}(2003)}]{matthias03}
\bibinfo{author}{\bibfnamefont{S.}~\bibnamefont{Matthias}} \bibnamefont{and}
  \bibinfo{author}{\bibfnamefont{F.}~\bibnamefont{M\"{u}ller}},
  \bibinfo{journal}{Nature} \textbf{\bibinfo{volume}{424}}, \bibinfo{pages}{53}
  (\bibinfo{year}{2003}).

\bibitem[{\citenamefont{Kettner et~al.}(2000)\citenamefont{Kettner, Reimann,
  H\"{a}nggi, and M\"{u}ller}}]{kettner00}
\bibinfo{author}{\bibfnamefont{C.}~\bibnamefont{Kettner}},
  \bibinfo{author}{\bibfnamefont{P.}~\bibnamefont{Reimann}},
  \bibinfo{author}{\bibfnamefont{P.}~\bibnamefont{H\"{a}nggi}},
  \bibnamefont{and}
  \bibinfo{author}{\bibfnamefont{F.}~\bibnamefont{M\"{u}ller}},
  \bibinfo{journal}{Phys. Rev. E} \textbf{\bibinfo{volume}{61}},
  \bibinfo{pages}{312} (\bibinfo{year}{2000}).

\bibitem[{\citenamefont{Gorre-Talini et~al.}(1998)\citenamefont{Gorre-Talini,
  Spatz, and Silberzan}}]{gorretalini98}
\bibinfo{author}{\bibfnamefont{L.}~\bibnamefont{Gorre-Talini}},
  \bibinfo{author}{\bibfnamefont{J.~P.} \bibnamefont{Spatz}}, \bibnamefont{and}
  \bibinfo{author}{\bibfnamefont{P.}~\bibnamefont{Silberzan}},
  \bibinfo{journal}{Chaos} \textbf{\bibinfo{volume}{8}}, \bibinfo{pages}{650}
  (\bibinfo{year}{1998}).

\bibitem[{\citenamefont{Bader et~al.}(1999)\citenamefont{Bader, Hammond, Henck,
  Deem, McDermott, Bustillo, Simpson, Mulhern, and Rothberg}}]{bader99}
\bibinfo{author}{\bibfnamefont{J.~S.} \bibnamefont{Bader}},
  \bibinfo{author}{\bibfnamefont{R.~W.} \bibnamefont{Hammond}},
  \bibinfo{author}{\bibfnamefont{S.~A.} \bibnamefont{Henck}},
  \bibinfo{author}{\bibfnamefont{M.~W.} \bibnamefont{Deem}},
  \bibinfo{author}{\bibfnamefont{G.~A.} \bibnamefont{McDermott}},
  \bibinfo{author}{\bibfnamefont{J.~M.} \bibnamefont{Bustillo}},
  \bibinfo{author}{\bibfnamefont{J.~W.} \bibnamefont{Simpson}},
  \bibinfo{author}{\bibfnamefont{G.~T.} \bibnamefont{Mulhern}},
  \bibnamefont{and} \bibinfo{author}{\bibfnamefont{J.~M.}
  \bibnamefont{Rothberg}}, \bibinfo{journal}{Proc. Nat. Acad. Sci.}
  \textbf{\bibinfo{volume}{96}}, \bibinfo{pages}{13165} (\bibinfo{year}{1999}).

\bibitem[{\citenamefont{Bader et~al.}(2002)\citenamefont{Bader, Deem, Hammond,
  Henck, Simpson, and Rothberg}}]{bader02}
\bibinfo{author}{\bibfnamefont{J.~S.} \bibnamefont{Bader}},
  \bibinfo{author}{\bibfnamefont{M.~W.} \bibnamefont{Deem}},
  \bibinfo{author}{\bibfnamefont{R.~W.} \bibnamefont{Hammond}},
  \bibinfo{author}{\bibfnamefont{S.~A.} \bibnamefont{Henck}},
  \bibinfo{author}{\bibfnamefont{J.~W.} \bibnamefont{Simpson}},
  \bibnamefont{and} \bibinfo{author}{\bibfnamefont{J.~M.}
  \bibnamefont{Rothberg}}, \bibinfo{journal}{Appl. Phys. A}
  \textbf{\bibinfo{volume}{75}}, \bibinfo{pages}{275} (\bibinfo{year}{2002}).

\bibitem[{\citenamefont{Faucheux
  et~al.}(1995{\natexlab{a}})\citenamefont{Faucheux, Bourdieu, Kaplan, and
  Libchaber}}]{faucheux95}
\bibinfo{author}{\bibfnamefont{L.~P.} \bibnamefont{Faucheux}},
  \bibinfo{author}{\bibfnamefont{L.~S.} \bibnamefont{Bourdieu}},
  \bibinfo{author}{\bibfnamefont{P.~D.} \bibnamefont{Kaplan}},
  \bibnamefont{and} \bibinfo{author}{\bibfnamefont{A.~J.}
  \bibnamefont{Libchaber}}, \bibinfo{journal}{Phys. Rev. Lett.}
  \textbf{\bibinfo{volume}{74}}, \bibinfo{pages}{1504}
  (\bibinfo{year}{1995}{\natexlab{a}}).

\bibitem[{\citenamefont{Faucheux
  et~al.}(1995{\natexlab{b}})\citenamefont{Faucheux, Stolovitzky, and
  Libchaber}}]{faucheux95a}
\bibinfo{author}{\bibfnamefont{L.~P.} \bibnamefont{Faucheux}},
  \bibinfo{author}{\bibfnamefont{G.}~\bibnamefont{Stolovitzky}},
  \bibnamefont{and}
  \bibinfo{author}{\bibfnamefont{A.}~\bibnamefont{Libchaber}},
  \bibinfo{journal}{Phys. Rev. E} \textbf{\bibinfo{volume}{51}},
  \bibinfo{pages}{5239} (\bibinfo{year}{1995}{\natexlab{b}}).

\bibitem[{\citenamefont{Koss and Grier}(2003)}]{koss03}
\bibinfo{author}{\bibfnamefont{B.~A.} \bibnamefont{Koss}} \bibnamefont{and}
  \bibinfo{author}{\bibfnamefont{D.~G.} \bibnamefont{Grier}},
  \bibinfo{journal}{Appl. Phys. Lett.} \textbf{\bibinfo{volume}{82}},
  \bibinfo{pages}{3985} (\bibinfo{year}{2003}).

\bibitem[{\citenamefont{Risken}(1989)}]{risken89}
\bibinfo{author}{\bibfnamefont{H.}~\bibnamefont{Risken}},
  \emph{\bibinfo{title}{The Fokker-Planck Equation}}
  (\bibinfo{publisher}{Springer-Verlag}, \bibinfo{address}{Berlin},
  \bibinfo{year}{1989}), \bibinfo{edition}{2nd} ed.

\bibitem[{\citenamefont{Igasaki et~al.}(1999)\citenamefont{Igasaki, Li,
  Yoshida, Toyoda, Inoue, Mukohzaka, Kobayashi, and Hara}}]{igasaki99}
\bibinfo{author}{\bibfnamefont{Y.}~\bibnamefont{Igasaki}},
  \bibinfo{author}{\bibfnamefont{F.}~\bibnamefont{Li}},
  \bibinfo{author}{\bibfnamefont{N.}~\bibnamefont{Yoshida}},
  \bibinfo{author}{\bibfnamefont{H.}~\bibnamefont{Toyoda}},
  \bibinfo{author}{\bibfnamefont{T.}~\bibnamefont{Inoue}},
  \bibinfo{author}{\bibfnamefont{N.}~\bibnamefont{Mukohzaka}},
  \bibinfo{author}{\bibfnamefont{Y.}~\bibnamefont{Kobayashi}},
  \bibnamefont{and} \bibinfo{author}{\bibfnamefont{T.}~\bibnamefont{Hara}},
  \bibinfo{journal}{Opt. Rev.} \textbf{\bibinfo{volume}{6}},
  \bibinfo{pages}{339} (\bibinfo{year}{1999}).

\bibitem[{\citenamefont{Crocker and Grier}(1996)}]{crocker96}
\bibinfo{author}{\bibfnamefont{J.~C.} \bibnamefont{Crocker}} \bibnamefont{and}
  \bibinfo{author}{\bibfnamefont{D.~G.} \bibnamefont{Grier}},
  \bibinfo{journal}{J. Colloid Interface Sci.} \textbf{\bibinfo{volume}{179}},
  \bibinfo{pages}{298} (\bibinfo{year}{1996}).

\bibitem[{\citenamefont{Behrens et~al.}(2003)\citenamefont{Behrens, Plewa, and
  Grier}}]{behrens03}
\bibinfo{author}{\bibfnamefont{S.~H.} \bibnamefont{Behrens}},
  \bibinfo{author}{\bibfnamefont{J.}~\bibnamefont{Plewa}}, \bibnamefont{and}
  \bibinfo{author}{\bibfnamefont{D.~G.} \bibnamefont{Grier}},
  \bibinfo{journal}{Euro. Phys. J. E} \textbf{\bibinfo{volume}{10}},
  \bibinfo{pages}{115} (\bibinfo{year}{2003}).

\bibitem[{\citenamefont{Ladavac et~al.}(2004)\citenamefont{Ladavac, Kasza, and
  Grier}}]{ladavac04}
\bibinfo{author}{\bibfnamefont{K.}~\bibnamefont{Ladavac}},
  \bibinfo{author}{\bibfnamefont{K.}~\bibnamefont{Kasza}}, \bibnamefont{and}
  \bibinfo{author}{\bibfnamefont{D.~G.} \bibnamefont{Grier}},
  \bibinfo{journal}{Phys. Rev. E} \textbf{\bibinfo{volume}{70}},
  \bibinfo{pages}{010901} (\bibinfo{year}{2004}).

\bibitem[{\citenamefont{Pelton et~al.}(2004)\citenamefont{Pelton, Ladavac, and
  Grier}}]{pelton04a}
\bibinfo{author}{\bibfnamefont{M.}~\bibnamefont{Pelton}},
  \bibinfo{author}{\bibfnamefont{K.}~\bibnamefont{Ladavac}}, \bibnamefont{and}
  \bibinfo{author}{\bibfnamefont{D.~G.} \bibnamefont{Grier}},
  \bibinfo{journal}{Phys. Rev. E} p. \bibinfo{pages}{accepted for publication}
  (\bibinfo{year}{2004}).

\bibitem[{\citenamefont{Hasson and Cheney}(2001)}]{hasson01}
\bibinfo{author}{\bibfnamefont{T.}~\bibnamefont{Hasson}} \bibnamefont{and}
  \bibinfo{author}{\bibfnamefont{R.~E.} \bibnamefont{Cheney}},
  \bibinfo{journal}{Curr. Opin. Cell Bio.} \textbf{\bibinfo{volume}{13}},
  \bibinfo{pages}{29} (\bibinfo{year}{2001}).

\bibitem[{\citenamefont{Lister et~al.}(2004)\citenamefont{Lister, Schmitz,
  Walker, Trinick, Buss, Veigel, and Kendrick-Jones}}]{lister04}
\bibinfo{author}{\bibfnamefont{I.}~\bibnamefont{Lister}},
  \bibinfo{author}{\bibfnamefont{S.}~\bibnamefont{Schmitz}},
  \bibinfo{author}{\bibfnamefont{M.}~\bibnamefont{Walker}},
  \bibinfo{author}{\bibfnamefont{J.}~\bibnamefont{Trinick}},
  \bibinfo{author}{\bibfnamefont{F.}~\bibnamefont{Buss}},
  \bibinfo{author}{\bibfnamefont{C.}~\bibnamefont{Veigel}}, \bibnamefont{and}
  \bibinfo{author}{\bibfnamefont{J.}~\bibnamefont{Kendrick-Jones}},
  \bibinfo{journal}{EMBO J.} \textbf{\bibinfo{volume}{23}},
  \bibinfo{pages}{1729} (\bibinfo{year}{2004}).

\bibitem[{\citenamefont{Okten et~al.}(2004)\citenamefont{Okten, Churchman,
  Rock, and Spudich}}]{okten04}
\bibinfo{author}{\bibfnamefont{Z.}~\bibnamefont{Okten}},
  \bibinfo{author}{\bibfnamefont{L.~S.} \bibnamefont{Churchman}},
  \bibinfo{author}{\bibfnamefont{R.~S.} \bibnamefont{Rock}}, \bibnamefont{and}
  \bibinfo{author}{\bibfnamefont{J.~A.} \bibnamefont{Spudich}},
  \bibinfo{journal}{Nature Struct. Mol. Bio.} \textbf{\bibinfo{volume}{11}},
  \bibinfo{pages}{884} (\bibinfo{year}{2004}).

\end{thebibliography}

\end{document}